\UseRawInputEncoding
\documentclass[
preprint,
 amsmath,amssymb,
 aps,
]{revtex4-1}

\usepackage{graphicx}
\usepackage{dcolumn}
\usepackage{bm}
\usepackage{siunitx}
\RequirePackage{lineno}
\DeclareSIUnit{\molar}{M}
\usepackage{todonotes}
\usepackage{booktabs}
\usepackage{physics}
\usepackage{mhchem}
\usepackage{float}
\usepackage{xcolor, soul} 
\usepackage{amsmath}

\begin{document}



\title{First passage time study of DNA strand displacement}

\author{DWB Broadwater, Jr.}
\thanks{These two authors contributed equally.}
\author{AW Cook}
\thanks{These two authors contributed equally.}
\author{HD Kim}

\date{\today}

\begin{abstract}
DNA strand displacement, where a single-stranded nucleic acid invades a DNA duplex, is pervasive in genomic processes and DNA engineering applications. The kinetics of strand displacement have been studied in bulk; however, the kinetics of the underlying strand exchange were obfuscated by a slow bimolecular association step. Here, we use a novel single-molecule Fluorescence Resonance Energy Transfer (smFRET) approach termed the ``fission" assay to obtain the full distribution of first passage times of unimolecular strand displacement. At a frame time of \SI{4.4}{\ms}, the first passage time distribution for a 14-nt displacement domain exhibited a nearly monotonic decay with little delay. Among the eight different sequences we tested, the mean displacement time was on average \SI{30}{\ms} and varied by up to a factor of 13. The measured displacement kinetics also varied between complementary invaders and between RNA and DNA invaders of the same base sequence except for T$\rightarrow$U substitution. However, displacement times were largely insensitive to the monovalent salt concentration in the range of \SIrange{0.25}{1}{\molar}. Using a one-dimensional random walk model, we infer that the single-step displacement time is in the range of \SIrange{\sim30}{\sim300}{\us} depending on the base identity. The framework presented here is broadly applicable to the kinetic analysis of multistep processes investigated at the single-molecule level.
\end{abstract}

\maketitle


\section{Statement of Significance}
DNA strand displacement occurs when a single nucleic acid strand invades and replaces another nearly identical strand in a duplex. This process is ubiquitous in biology and is fundamental to the field of DNA nanotechnology. Previous kinetic studies of strand displacement either used DNA strands much longer than those found in practical applications or were obscured by a rate-limiting bimolecular step known as toehold formation. In this study, we introduce a new, single-molecule scheme that enables direct measurement of the strand displacement first passage time. Our observed kinetics demonstrate highly non-trivial sequence dependence as well as surprising differences between RNA and DNA invaders.

\section{Introduction}
Nucleic acids' ability to form hydrogen bonds between complementary Watson-Crick bases allows for a rich set of complicated, multi-step kinetic behaviors such as duplex hybridization\cite{ouldridge2013dna} and dehybridization\cite{sanstead2018direct}, Holliday junction structural dynamics\cite{mckinney2003structural,bugreev2006rad54}, and strand invasion\cite{wright2018homologous}. In particular, strand displacement, which is the exchange of bases between two competing nucleic acid strands of identical sequence, occurs in homologous recombination\cite{chen2008mechanism,savir2010reca}, DNA replication\cite{mi2020strand} and RNA transcription\cite{kireeva2018rna}, as well as CRISPR/Cas\cite{singh2016real} and the related Cascade complex\cite{ivanvcic2012tuning}. In addition to fundamental genomic processes, DNA nanotechnology exploits strand displacement to create nanoscale gadgets\cite{andersen2009self,thubagere2017cargo, chang2019simply} and computational circuits\cite{zhang2011dynamic, cherry2018scaling, wang2020implementing,simmel2019principles}. Strand displacement also aids in the development of quantitative assays for detection of nucleic acid\cite{wang2018autonomous,li2019cellular,sapkota2019single} and enzymatic activity\cite{cui2019integrated, lee2019fluorescent} with improved probe specificity\cite{figg2020controlling,tang2020dna,garcia2020stability}. 

\begin{figure}[th!]
\centering
    \includegraphics[width=\linewidth,keepaspectratio]{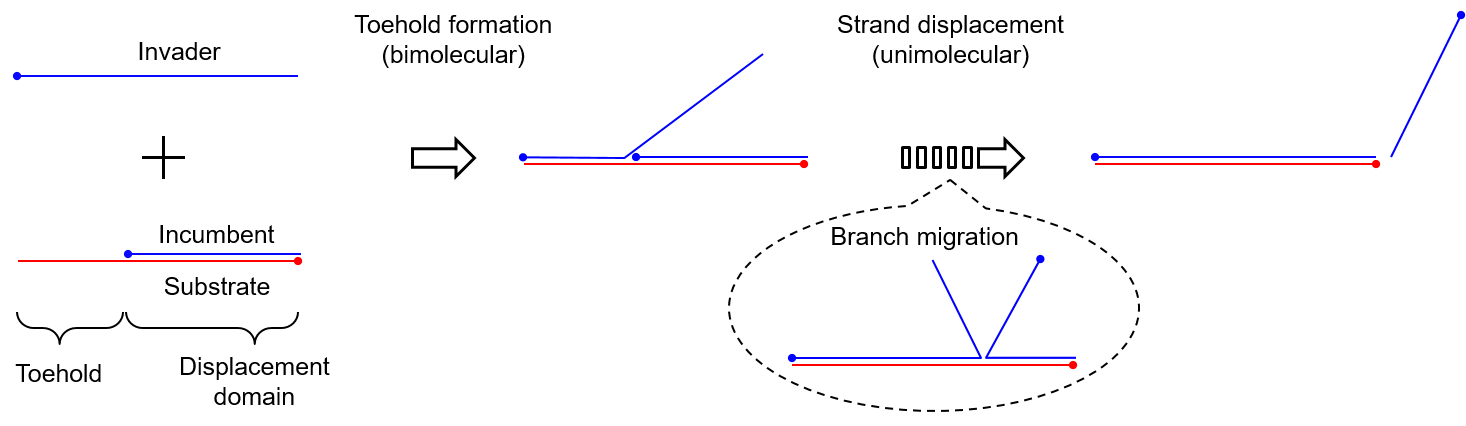}
    \caption{Toehold-mediated DNA strand displacement. Each line represents a single-stranded DNA with the $5^\prime$-end marked by a dot. Blue and red colors represent complementary sequences. In bulk studies, the apparent reaction kinetics are dominated by the slow bimolecular step. In this study, we focus on the unimolecular strand displacement, which is thought to occur through branch migration.}
    \label{fig:reaction}
\end{figure}

For practical applications, strand displacement is implemented with the ``invader" strand and a partial duplex composed of the ``incumbent" strand and the ``substrate" strand\cite{simmel2019principles}(Fig.~\ref{fig:reaction}). The partial duplex has two distinct domains: 1) the single-stranded overhang called the toehold which is critical to the speed and efficiency of the reaction\cite{zhang2009control} and 2) the duplex region called the displacement domain. Toehold-mediated strand displacement is initiated when the invader strand anneals to the toehold in a bimolecular reaction. Once a stable toehold interaction is formed, the incumbent can be displaced by the dangling strand of the invader through spontaneous opening of a base pair between substrate and incumbent and closing of a base pair between substrate and invader. This unimolecular strand displacement is also called branch migration\cite{radding1977uptake,green1981reassociation,srinivas2013biophysics}. 

Great attention has been paid to the kinetics of toehold-mediated strand displacement \cite{zhang2009control,qian2011scaling,srinivas2013biophysics,machinek2014programmable,broadwater2016effect}. However, these kinetics have been measured mostly in bulk, where the reaction kinetics are limited by bimolecular toehold association. Therefore, the measured kinetics do not shed light on the unimolecular branch migration. Other studies using long (\SI{\sim1}{kbp}) DNA estimated the branch migration time per base pair step to be \SI{\sim 10}{\us}\cite{radding1977uptake,green1981reassociation}, but strand displacement in those studies took place in a D-loop geometry, which is different from the geometry of current interest where dangling strands are unrestricted during branch migration. Also, \SI{\sim1}{kbp} scale extends far beyond the length scales of interest for DNA nanotechnology. To understand whether and how sequence can be used to control displacement kinetics, we require experimental studies on unimolecular displacement of short oligos that can be modeled at the single base level.

In this study, we introduce a DNA ``fission" assay to study toehold-mediated DNA strand displacement kinetics. The fission assay employs single-molecule Fluorescence Resonance Energy Transfer (smFRET) in order to directly measure the first passage displacement time\cite{chou2014first,polizzi2016mean} for the unimolecular reaction that occurs between toehold formation and incumbent dissociation. Using a wide-field total internal reflection fluorescence microscope, we measured the displacement kinetics for a 14-nucleotide displacement domain of eight different sequences. The mean displacement time varied by more than 10-fold between the slowest and fastest sequence and was on average \SI{\sim30}{\ms}, and the histograms of displacement times obtained at \SI{4.4}{ms} resolution showed a monotonous decay with little to no lag. We found that the displacement kinetics depend on the base sequence and the nucleic acid type (DNA vs. RNA) of the invader, but not on monovalent salt concentration. We analyzed the first passage time histograms of strand displacement using a symmetric random walk model to extract single base pair step times. The best fit to the histograms was obtained with \SI{\sim 33}{\us}, \SI{\sim 200}{\us}, \SI{\sim 250}{\us}, and \SI{\le 33}{\us} for A,C,T, and G respectively. Our study reports the displacement rates of short DNA oligos and reveals biophysical mechanisms that govern DNA strand displacement kinetics.  

\section{Materials and Methods}

\subsection{Sample Preparation}
Custom DNA oligomers were purchased from Integrated DNA Technologies. The 26-nt substrate was internally labeled near the end distal to the toehold with a Cy3 fluorophore. The 24-nt invader molecule was labeled with a BioTEG linker at the end proximal to the toehold for surface immobilization. The 14-nt incumbent sequences were labeled with a Cy5 fluorophore at the end distal to the toehold. All oligos were HPLC purified by the manufacturer. The specific sequences are in Tables \ref{table:invader}$-$\ref{table:substrate} in the Supporting Material. Partial duplexes were constructed by combining substrate and incumbent at a 1:10 ratio ($\SI{0.5}{\micro \molar}$ substrate, $\SI{5.0}{\micro \molar}$ incumbent) in buffer at 7 pH containing $\SI{1}{\molar}$ NaCl and $\SI{10}{\milli \molar}$ Tris. The excess of incumbent strands was meant to minimize the number of single-stranded substrates in solution; unpaired substrates can compete with the partial duplexes for binding with the surface-bound invaders, while lone incumbent strands do not bind to the invaders and will not fluoresce on their own. The mixture was heated to $\SI{95}{\celsius}$ and slowly cooled for \SI{3}{\hour} to $\SI{4}{\celsius}$ to ensure the partial duplex was fully annealed.

\subsection{Experimental Setup}
Molecules were observed with an objective-type total internal reflection fluorescence microscope assembled on a commercial microscope body (IX81; Olympus). Fluorophores were excited by a $\SI{532}{nm}$ laser (BWN-532-50E, B\&W Tek). Images were $2\times2$ binned and captured with an EMCCD (DU-897ECS0-\#BV; Andor Technology), and images were recorded at 228 fps with $\SI{3.96}{\milli \second}$ exposure time using Micro-Manager software\cite{edelstein2014advanced}. This high frame rate was achieved by cropping the image height to 64 super pixels. Experiments were performed on flow cells constructed as previously described in Le and Kim\cite{le2014studying} while flow volume and flow rate ($\SI{900}{\micro \liter \per \minute}$) was controlled by a syringe pump (NE-1000; New Era Pump System).

The surface was passivated with polyethylene glycol (PEG) to minimize nonspecific binding\cite{le2014studying}. After neutravidin coating, the biotin-containing invader molecules were immobilized by flowing in at a concentration of $\SI{1}{\nano\molar}$. Next, $\SI{20}{\micro\liter}$ of partial duplexes were pumped into the flow cell at $\SI{2.5}{\nano \molar}$ in an oxygen-scavenging imaging buffer\cite{aitken2008oxygen}, which contained $\SI{1}{\milli \molar}$ 6-hydroxy-2,5,7,8-tetramethylchroman-2-carboxylic acid (Trolox), $\SI{5}{\milli \molar}$ protocatechuic acid, $\SI{100}{\nano \molar}$ protocatechuate 3,4-dioxygenase, and $\SI{100}{\milli \molar}$ Tris-HCl (pH 7).

An appearance of high FRET signal marked formation of the toehold. A low FRET signal appeared as strand displacement concluded. The FRET signal time series was recorded and analyzed using in-house MATLAB software. The lifetime of the high FRET state was observed for many molecules to collect a distribution of displacement times. 

\subsection{Statistics of displacement times}
Here, we provide an analytical expression we used to fit the histograms of displacement times. We model strand displacement as a one-dimensional random walk: 
\begin{equation}
\ce{0 <-->[$f_0$][$r_1$] 
1
<-->[$f_1$][$r_2$]
2
<-->[$f_2$][$r_3$]
$\cdots$
<-->[$f_{n-1}$][$r_n$]
$n$
\label{eq:1d}
}
\end{equation}
In this model, each state is denoted by $i$, the number of displaced bases, and the measured displacement time corresponds to the first passage time from the reflecting state $0$ on the left boundary to the absorbing state $n$ on the right boundary. The forward rate and reverse rate from state $i$ are denoted as $f_i$ and $r_i$, respectively. Since state $n$ is the absorbing state, $r_n=0$. The time dependence of the system is governed by the master equation
\begin{equation}
    \label{master}
    \frac{\partial\ket{\psi(t)}}{\partial t}=\vectorbold{L}\ket{\psi(t)},
\end{equation}
where $\vectorbold{L}$ is the transition matrix operator, and the ket vector $\ket{\psi(t)}$ represents the system state. The probability amplitude to be in the absorbing state $\ket{n}$ at some time $t$ after starting in $\ket{0}$ is then given by\cite{hartich2019interlacing}
\begin{equation}
    \label{eq:prob1}
    P(n,t|0)=\bra{n}e^{\vectorbold{L} t}\ket{0}. 
\end{equation}
The experimentally accessible datapoints in single-molecule experiments are the number of displacement events $\Delta N$ detected during a short time interval or bin time $\Delta t$. These numbers form the so-called dwell-time or survival-time histogram. For sufficiently large number of total events $N_0$, $\Delta N$ in the $i$-th bin is related to the probability amplitudes according to
\begin{align}
\frac{\Delta N(i)}{N_0} &= P(n,i\Delta t|0)-P(n,(i-1)\Delta t|0) \nonumber \\ 
&= \bra{n}e^{\vectorbold{L} i\Delta t}\left( 1-e^{-\vectorbold{L}\Delta t}\right)\ket{0}.
\end{align}
This can be expanded using the left and right eigenvectors of $\vectorbold{L}$, $\bra{\phi_k^L}$ and $\ket{\phi_k^R}$ that satisfy $\braket{\phi_m^L}{\phi_n^R}=\delta_{mn}$, and their corresponding eigenvalue $-\mu_k$:
\begin{equation}
\frac{\Delta N(i)}{N_0} = \sum_k \braket{n}{\phi_k^R}\braket{\phi_k^L}{0} e^{-\mu_k i\Delta t} \left(1 - e^{\mu_k\Delta t}\right)
\label{eq:dwelltime}
\end{equation}
In the limit of $\Delta t\rightarrow 0$, Eq.~\ref{eq:dwelltime} yields the first passage time density $f(t)$
\begin{equation}
    \label{eq:fptd}
 f(t) \equiv \lim_{\Delta t\rightarrow0} \frac{1}{N_0}\frac{\Delta N(i)}{\Delta t}=\pdv{t}P(n,t|0) =\sum_k \braket{n}{\phi_k^R}\braket{\phi_k^L}{0} (-\mu_k) e^{-\mu_k t}.
\end{equation}
We used Eq.~\ref{eq:dwelltime} to fit the measured histograms of displacement times with a fixed $\Delta t$ that corresponds to the frame time of 4.4 ms. In the representation of $\ket{0}\mapsto(1,0,0,...,0)^\mathrm{T}$, $\vectorbold{L}$ is an asymmetric tridiagonal matrix:
\begin{equation}
\vectorbold{L} = \begin{bmatrix}
-f_0   &        r_1      &         0        & \cdots &                0                 &             0            \\
f_0      & -(f_1+r_1) &       r_2       &  \cdots      &             0               &          0          \\
0       &         f_1     & -(f_2+r_2) & \cdots &                0                 &             0            \\
\vdots       &         \vdots       &       \vdots       & \ddots &             \vdots             &             \vdots           \\
0      &      0      &      0            & \cdots  & -(f_{n-1}+r_{n-1}) &        0          \\
 0       &          0       &      0      &    \cdots  &             f_{n-1}             & 0 \\
\end{bmatrix} .
\end{equation}
whose left and right eigenvectors and eigenvalues can be obtained using MATLAB. 

We also present here the expression we use to analyze the mean first passage time $\tau$\cite{kim1958mean}. $\tau$ can be computed using Eq.~\ref{eq:fptd} as
\begin{equation}
\label{eq:mfptd}
\tau =\int_0^\infty t f(t) dt.
\end{equation} A more useful expression can be obtained in terms of an invertible submatrix of $\mathbf{L}$, which we term $\mathbf{A}$:
\begin{equation}
\mathbf{L}=
\left[
\begin{array}{c|c}
   \raisebox{-15pt}{{\huge{[$\mathbf{A}$]}}} & 0 \\[-4ex] 
   & \vdots \\[-0.5ex]
   & 0 \\ \hline
  0 \cdots f_{n-1} 
  & 0
\end{array}
\right].
\end{equation}
Using the normalization of probability amplitude 
\begin{equation}
    P(n,t|0)=1-\sum_{i=0}^{n-1}P(i,t|0) 
\end{equation}
we can express $\tau$ in terms of the inverse of $\mathbf{A}$
\begin{align}
\int_0^\infty tf(t)dt &=-\int_0^\infty t d\left[ \sum_{i=0}^{n-1}\bra{i}e^{-\mathbf{A} t}\ket{0} \right] \nonumber\\
&=\int_0^\infty \sum_{i=0}^{n-1}\bra{i}e^{-\mathbf{A} t}\ket{0} dt\nonumber\\
&=\sum_{i=0}^{n-1}\bra{i}\mathbf{A}^{-1}\ket{0}.
\label{eq:sum}
\end{align}
In the matrix presentation, the inverse matrix $\mathbf{A}^{-1}$ is related to matrix cofactors by 
\begin{equation}
    \mathbf A^{-1} = \frac{1}{\operatorname{det}(\mathbf A)} \mathbf C^\mathsf{T}
    \label{eq:inverse}
\end{equation}
Plugging Eq.~\ref{eq:inverse} into Eq.~\ref{eq:sum}, 
\begin{align}
    \tau=\frac{1}{\operatorname{det}(\mathbf A)}\sum_{j=1}^n C_{1j}.
\end{align}
This sum of cofactors can be equated to the determinant of matrix $\mathbf A'$ which replaces the first row of $\mathbf A$ with 1's. Hence, the mean first passage time is given by the ratio of two matrix determinants\cite{broadwater2016effect}:  
\begin{equation}
\tau=\frac{
\begin{vmatrix}
1 & 1 & \cdots &1\\
a_{21} & a_{22} & \cdots &a_{2n}\\
\vdots & \vdots & \ &\vdots\\
a_{n1} & a_{n2} & \cdots &a_{nn}
\end{vmatrix}}{
\begin{vmatrix}
a_{11} & a_{12} & \cdots &a_{1n}\\
a_{21} & a_{22} & \cdots &a_{2n}\\
\vdots & \vdots & \ &\vdots\\
a_{n1} & a_{n2} & \cdots &a_{nn}
\end{vmatrix}} ,
\label{eq:mfpt}
\end{equation}
where $a_{ij}$ represent the matrix elements of $\mathbf A$. Eq.~\ref{eq:mfpt} can also be expressed in terms of the bias factor $\alpha_i=f_{i-1}/r_i$ as\cite{broadwater2016effect}
\begin{equation}
    \tau=\frac{1}{f_0}+\frac{1+\alpha_1}{\alpha_1}\cdot\frac{1}{f_1}+\frac{1+\alpha_1+\alpha_2\alpha_1}{\alpha_2\alpha_1}\cdot\frac{1}{f_2}+\cdots.
    \label{eq:mfpt2}
\end{equation}

\begin{figure}[th!]
\centering
    \includegraphics[width=\linewidth,keepaspectratio]{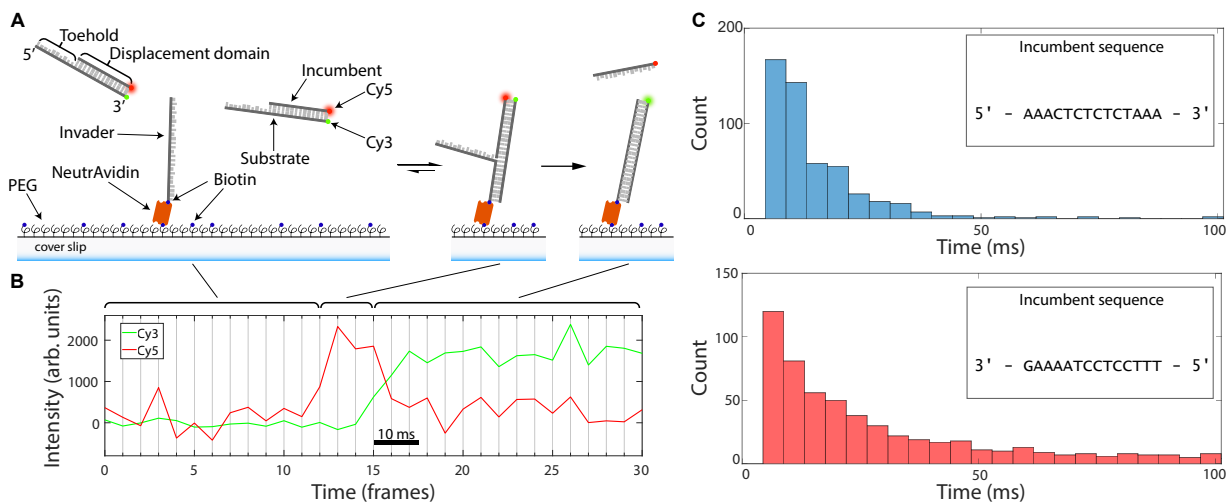}
    \caption{An overview of strand displacement data acquisition. \textbf{(A)} Experimental scheme of fission assay. Invader strands were immobilized on a PEG-passivated coverslip surface via a biotin-NeutrAvidin linker. Partial duplexes were labeled with a FRET pair (Cy5, incumbent; Cy3, substrate) and were drawn into the flow cell. After a diffusive search process, the partial duplex binds to the toehold, and the incumbent strand is displaced. \textbf{(B)} Sample acceptor and donor time traces. Acceptor (Cy5) signal increases upon toehold binding, and high FRET signal is sustained until displacement occurs which is indicated by a low FRET signal. The displacement time is identified as the high FRET lifetime. \textbf{(C)} Displacement time distributions for different sequences. Displacement time distributions show sequence dependence in the displacement domain (incumbent sequence shown). Displacement lifetimes are collected for many traces and assembled into a distribution. The binning size is the single frame time ($\SI{4.4}{\milli \second}$). }
    \label{fig:schematic}
\end{figure}

\section{Results}
To focus on the unimolecular kinetics of strand displacement, we took a surface-based single-molecule FRET approach (Fig.~\ref{fig:schematic}). In this approach, the invader is immobilized on the glass surface of a flow chamber, and the partial duplex between the donor (Cy3)-labeled substrate and the acceptor (Cy5)-labeled incumbent are perfused into the chamber. The toehold length (10-bp) is chosen so that toehold formation is practically irreversible throughout the experiment. Upon toehold formation, a diffraction-limited spot emerges out of the diffusive background in the Cy5 channel. Upon incumbent dissociation, the spot changes fluorescence emission from the Cy5 channel to the Cy3 channel. We termed this experimental scheme ``fission" because the duplex labeled with the FRET pair is split as a result of strand displacement.

Partial duplexes were constructed by annealing Cy3-labeled substrate molecules and Cy5-labeled incumbent molecules. Invader molecules were biotinylated near the end containing the toehold sequence and immobilized onto the surface (see Fig.~\ref{fig:schematic}(A)). As shown in Fig.~\ref{fig:schematic}(B) and SFig.~\ref{sfig:rawdata}, high FRET signals started to appear in the field of view after partial duplexes were flowed into the chamber. The average time at which spots appeared became shorter at a higher concentration of partial duplexes, and the transition of FRET from high to low only occurred in the presence of the matching displacement domain. Without the matching displacement domain, the high-FRET spots remained until they photobleached, which confirms that dissociation of the 10-bp toehold is much slower than the typical minute-long observation period. The red signal jumped to a high level in one or two frames, which suggests that toehold formation is much faster than our time resolution, and can therefore be considered instantaneous for analysis purposes. This high-level red signal lasted for variable periods of time from trace to trace, but the eventual transition back to low-FRET always occurred in one or two frames. Simultaneously with the disappearance of the signal from the Cy5 channel, a new signal appeared in the Cy3 channel, consistent with the fission scheme (Fig.~\ref{fig:schematic}(A)). Based on these observations, the first arrival of a high-FRET spot was attributed to toehold formation, and the transition from high- to low-FRET was attributed to completion of strand displacement. Hence, the dwell time in the high-FRET state (Fig.~\ref{fig:schematic}(B)) represents the displacement time.

\begin{figure}[th!]
\centering
    \includegraphics[width=12.5cm]{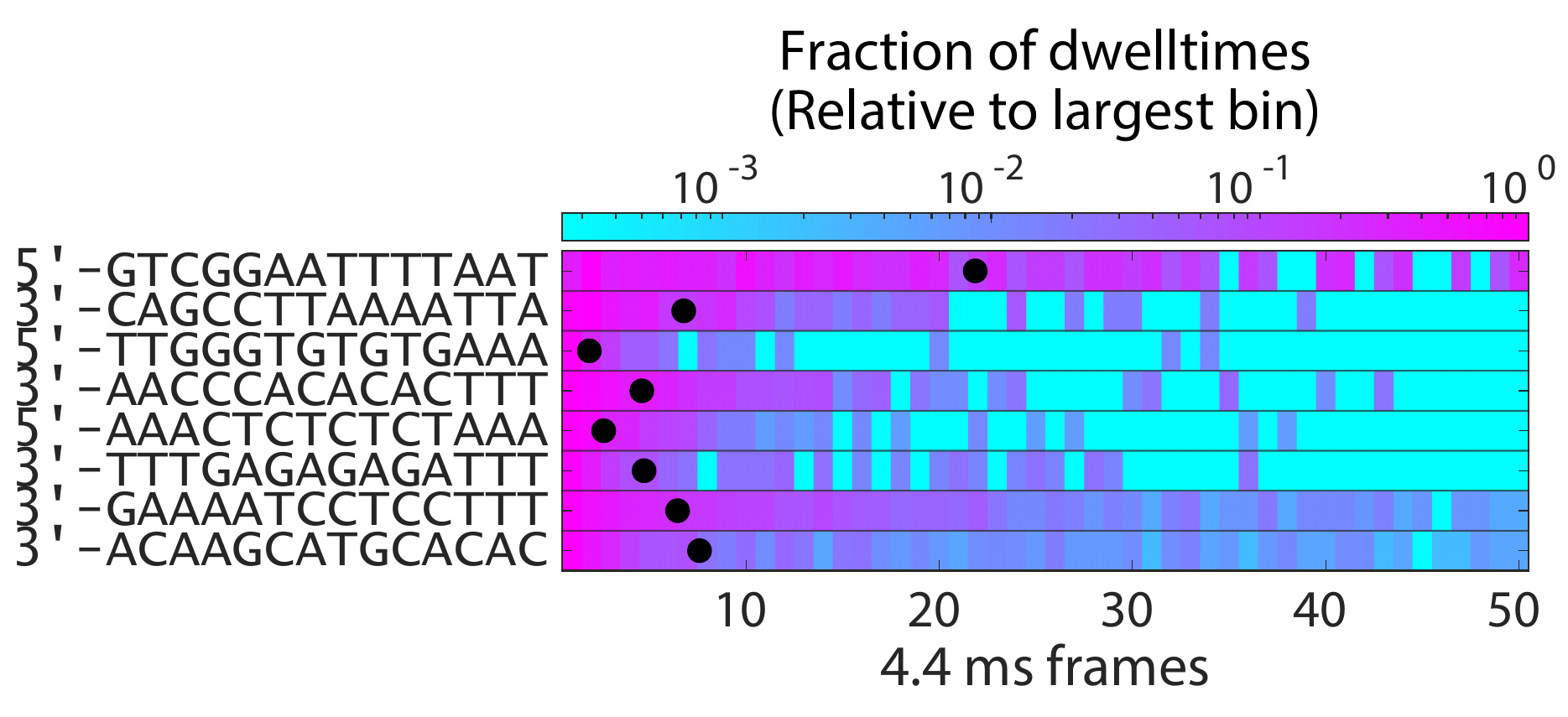}
    \caption{A heatmap demonstrating the distribution of displacement times within the first fifty frames of a typical experiment for each of the eight featured sequences. The sequences of the incumbent strands are listed to the left. The colormap is set to a logarithmic scale in order to better represent the exponentially decaying distributions. Here, each experiment is scaled such that the most populous bin is set to unity. The second bin is the most populous in experiments with incumbent strands 5'-GTCGGAATTTTAAT and 3'-CAGCCTTAAAATTA; in all other experiments, the first bin is the most populous. The mean of all recorded displacement times is marked by a black dot for each experiment. From top to bottom, the number of events recorded for each experiment is: 273, 751, 143, 890, 522, 169, 627, 609.}
    \label{fig:heatmap}
\end{figure}

By performing the fission assay multiple times, we could record hundreds of strand displacement events for one particular displacement system and build a histogram of displacement times. To investigate the sequence dependence of strand displacement kinetics, we tested 8 unique strand displacement systems, each with a different sequence in the displacement domain. We obtained these histograms at the finest bin width of $4.4$ ms, two of which are shown in Fig.~\ref{fig:schematic}(C). Note that displacement events faster than the exposure time do not produce a clear signal in the acceptor channel, and therefore, the first bin of the histogram starts from $4.4$ ms. For comparison of the histogram across all eight different sequences, we also present the histograms as a normalized heat map in Fig.~\ref{fig:heatmap}. The salient feature of these histograms is that they decay monotonically with little or no delay. Six out of eight sequences show decay from the first bin; only two sequences show more events in the second than in the first bin. Nonetheless, we find a significant difference in the characteristic decay time among the tested sequences (black dots, Fig.~\ref{fig:heatmap}). The fastest mean displacement time is 8 ms, while the slowest is 107 ms. The average over all sequences is \SI{35}{\ms}. 

\begin{figure}[th!]
\centering
    \includegraphics[width=8.5cm]{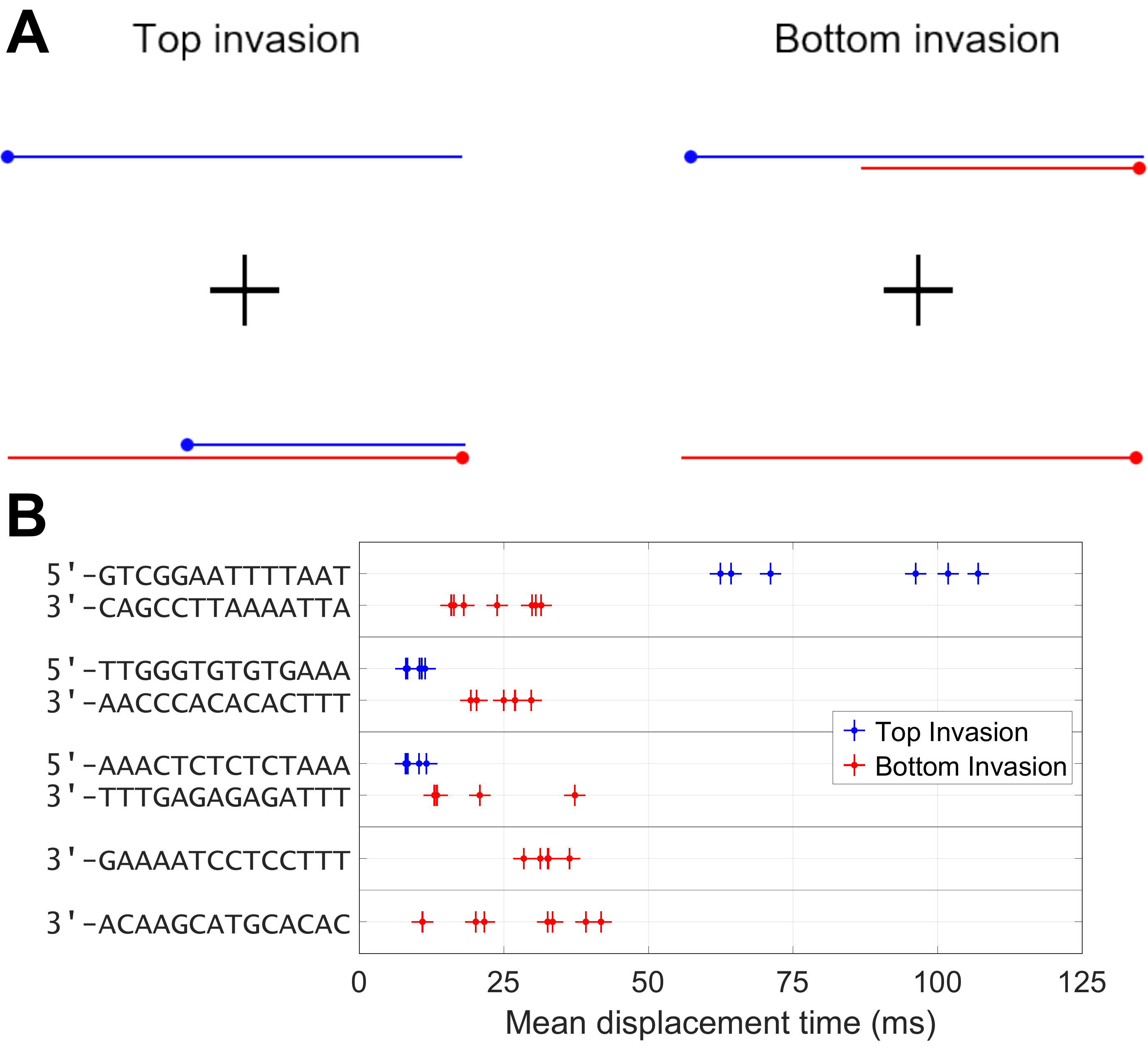}
    \caption{Kinetics of various displacement systems. \textbf{(A)} Schematic of invasion designs. Top invasion is defined by an invader with a 5$'$ toehold. In bottom invasion, all strands are replaced by their reverse complement. Bottom invasion is defined by an invader with a 3$'$ toehold. These systems are highly related as duplex base pairing is identical in both systems. \textbf{(B)} Mean displacement times. The mean displacement time is calculated from data below the 95\textsuperscript{th} percentile. The data collection process was entirely repeated several times for each invasion system. The mean times can change by an order of magnitude depending on the sequence. The sequences are the invader nucleotides that are beyond the toehold with the nucleotide most proximal to the toehold written first. The top invasion sequences therefore correspond to the conventional $5'$ to $3'$ direction, while the bottom invasion sequences are listed $3'$ to $5'$.}
    \label{fig:meantimes}
\end{figure}

To ensure that the observed difference between different sequences is not due to the uncertainties of the histograms, we need to establish the baseline uncertainties in the empirical histograms. As explained above, each histogram is obtained by combining displacement events taken from multiple runs of the fission assay in one day using the same reagents and flow cell. Hence, each histogram possesses statistical uncertainty due to the finite number of events and empirical uncertainty due to fluctuations in the experimental conditions. To estimate these uncertainties, we randomly sampled $80\%$ of the events collected on the same day and re-evaluated the mean displacement time (Supp. Fig.~\ref{sfig:mean sampling}). The spread of the mean values is nonuniform among different sequences. For example, Sequences 4 and 6 have a similar total mean, but show different uncertainties. Nonetheless, the uncertainty in the mean for each sequence is much narrower than the variation among different sequences. We also documented the variability of the histogram means obtained at different times over a 4-year span by two users (Supp. Fig.~\ref{sfig:means old new}). This empirical variability is much higher than the statistical variability due to reasons not completely clear. For transparency, we present these individual mean values in Fig.~\ref{fig:meantimes}.    

\begin{figure}[th!]
\centering
    \includegraphics[width=12.5cm]{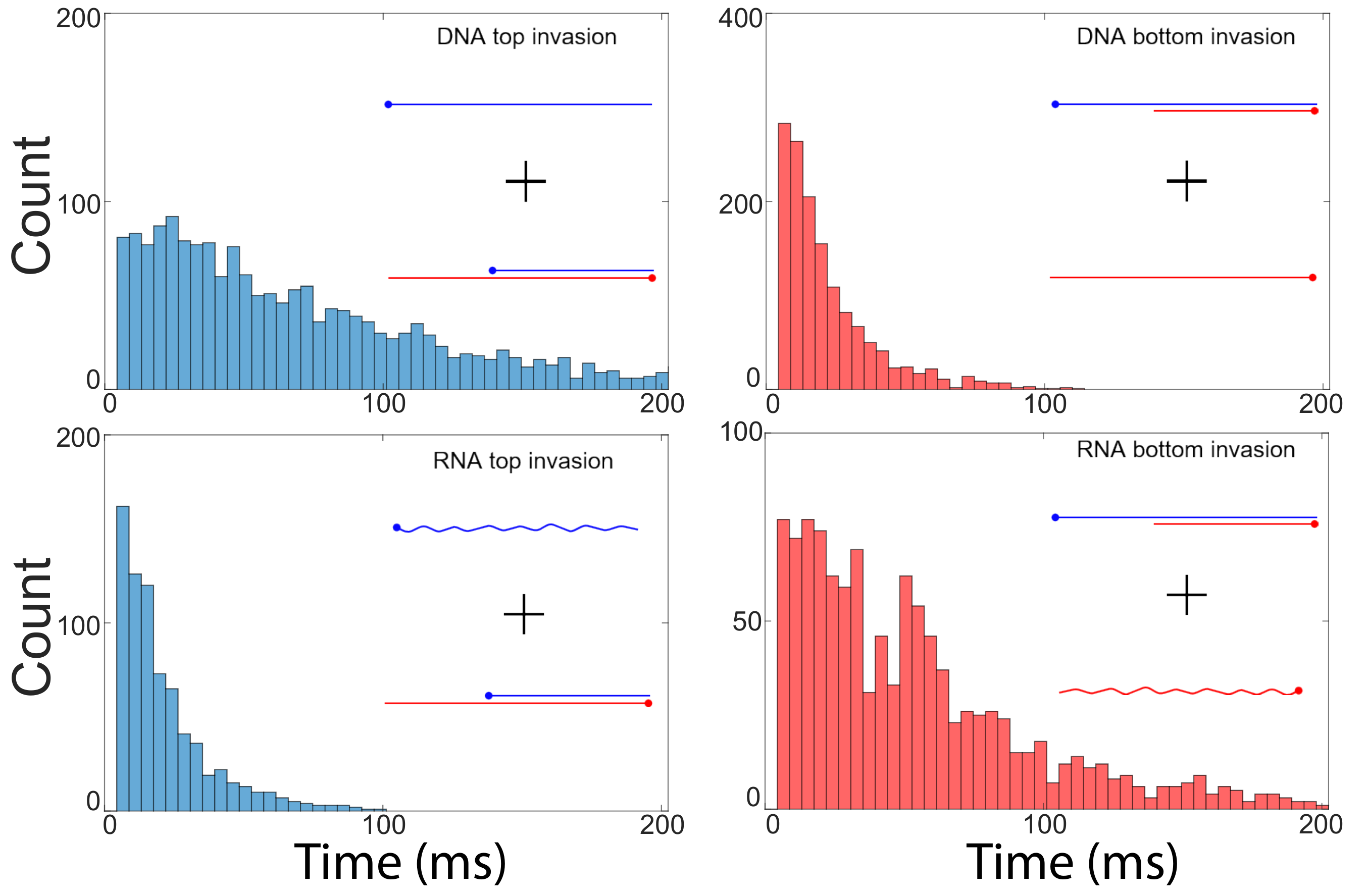}
    \caption{Comparison between RNA and DNA invasion. We measured displacement times for a pair of complementary DNA invaders which exhibited the largest difference between sides of invasion (blue, top invasion; red, bottom invasion) in comparison to all other pairs of invaders we observed (top row, straight invader). We measured displacement times for RNA versions of the invaders which were identical in sequence except for a T$\rightarrow$U substitution (bottom row, wavy invader). DNA partial duplexes were employed in all cases. Again, we found a difference in passage time depending on the side of invasion. Further, we noticed that the relative times switched with respect to the side of invasion.}
    \label{fig:RNA}
\end{figure}

In addition to the base pair sequence of the displacement domain, the base sequence of the invader can also affect the displacement kinetics. As shown in Fig.~\ref{fig:meantimes}(A), the same displacement domain can be invaded using a toehold extended from either the $5^\prime$-end or the $3^\prime$-end of the displacement domain. We refer to these complementary invasions as top and bottom invasions. The mean displacements times of top and bottom invasions are clearly different for all three displacement domains we tested. No particular invasion side was consistently faster: for Sequence 1, bottom invasion is faster, but for Sequence 2, top invasion is. Interestingly, when RNA with an identical sequence except for T$\rightarrow$U substitution was used as an invader in place of DNA, the faster side was switched (Fig.~\ref{fig:RNA}). All of these results suggest that the displacement kinetics are not completely determined by the base pair sequence or the thermodynamic stability of the displacement domain, but rather that the measured displacement kinetics are sensitive down to the chemical makeup of invading bases. 

\begin{figure}[th!]
\centering
    \includegraphics[width=8.5cm]{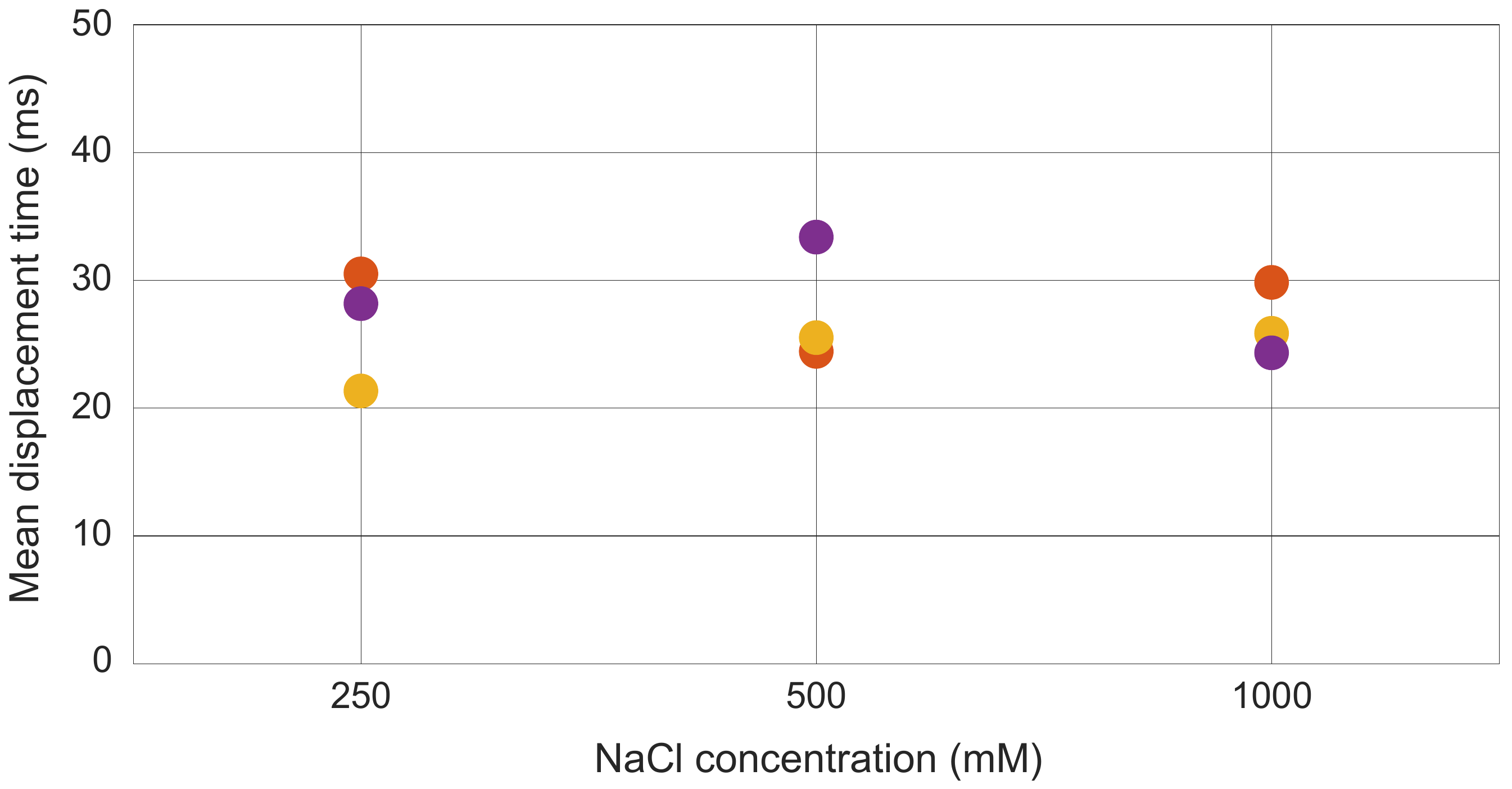}
    \caption{Salt dependence of mean displacement time. Data points sharing the same color were collected on the same day using the same reagents, aside from partial duplex solutions containing different concentrations of NaCl. The mean displacement time does not appear to depend strongly on the NaCl concentration. The incumbent sequence used for these measurements is 5$'$-ATTAAAATTCCGAC-3$'$ }
    \label{fig:salt}
\end{figure}

Lastly, we investigated the salt dependence of displacement kinetics. Monovalent salt can screen the negative charges on the phosphate backbone and alter the thermodynamics and kinetics of base pairing\cite{huguet2010single}. However, its effect on the kinetics of branch migration is less clear because branch migration involves the base pairing dynamics of two competing strands. As shown in Fig.~\ref{fig:salt}, the mean displacement time shows little change from 250 mM to 1 M [NaCl].   

\section{Discussion}
Using the fission assay, we measured the unimolecular branch migration kinetics in toehold-mediated DNA strand displacement. Using wide-field TIRFM and subregion readout of an EMCCD camera, we were able to record tens of strand displacement events at 4.4 ms frame rate. Our fission assay begins in a dark field-of-view with unlabeled invader strands immobilized on the surface, and monitors displacement events through the appearance and disappearance of FRET signal on the surface. The experimental design permits us to use high excitation intensity to detect fast displacement events at high signal-to-noise; strong excitation of fluorescent molecules begins only at the start of branch migration. Hence, the undesirable effect of photobleaching is eliminated. 

Our fission assay produces data that could not be obtained to date. It separates out the bimolecular toehold formation step from the rest so that the apparent displacement time truly reflects a unimolecular process. In the language of stochastic processes, the displacement time represents the first passage time: the time taken for the branch point to start from the first position and reach the last for the first time. The fission scheme allows access to the full distribution of individual displacement times, which is more informative than just the average values. Below, we use the first passage time analysis to extract single-step migration rates from the measured histograms and discuss potential microscopic mechanisms that may control these rates.

\begin{figure}[th!]
\centering
    \includegraphics[width=12.5cm]{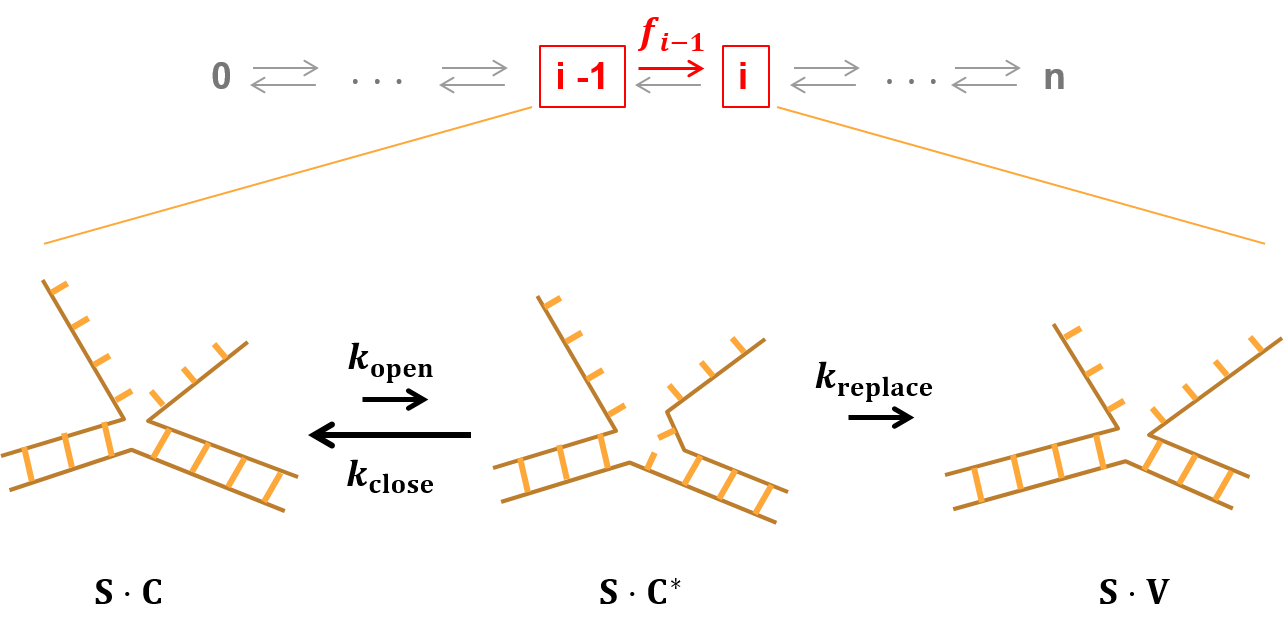}
    \caption{A closer look at the single base pair step transition. Any forward arrow in the one-dimensional Markov chain involves opening of a base pair between the substrate (S) and the incumbent (C) and closing of a base pair between the substrate and the invader (V). For a reverse arrow, the roles of incumbent and invading bases are simply flipped. Hence, each step can be modeled as a first passage from an initial state S$\cdot$C to a final state S$\cdot$V through an intermediate state S$\cdot$C* where a single base pair is transiently open. The transiently unbound incumbent base can rebind the substrate base at a rate of $k_\mathrm{close}$ or be replaced with the invading base at a rate of $k_\mathrm{replace}$.}. 
    \label{fig:singlestep}
\end{figure}
The most elementary model to describe DNA strand displacement is a one-dimensional random walk among states defined by the number of displaced base pairs (Eq.~\ref{eq:1d}). Displacement is initiated after the invader hybridizes to the toehold and continues until the incumbent loses all base pairs with the substrate to the invader. Any intermediate state during this process can be envisioned as two dangling strands branching off from the duplex stem (Fig.~\ref{fig:singlestep}). At the junction or the branch point, an incumbent (invader) base can spontaneously break away from the substrate base, and the most adjacent invader (incumbent) base can base-pair with the substrate base. As a result, the branch point can move by one base in either direction. The branch point, however, cannot recede into the toehold region because the incumbent is shorter than the substrate. Therefore, branch migration can be modeled as a one-dimensional random walk with single base steps from a reflecting boundary on one end (state 0) to an absorbing boundary on the other (state n).

\begin{figure}[th!]
\centering
    \includegraphics[width=12.5cm]{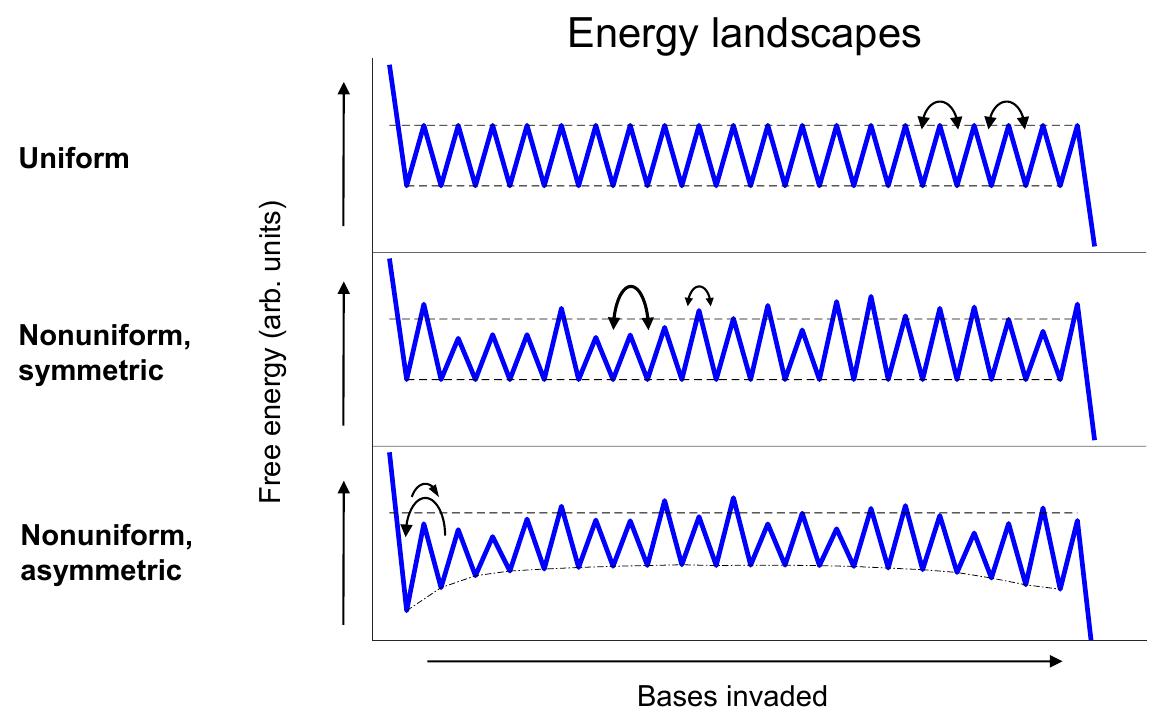}
    \caption{Free energy landscapes. All landscapes contain a reflecting boundary to the left and an absorbing boundary to the right. The uniform landscape is characterized by one uniform rate for transitions in any direction. The nonuniform, symmetric landscape allows for variation in rates so long as the forward rate from one state to another is set equal to the corresponding reverse rate. The nonuniform, asymmetric landscape additionally allows forward and reverse rates to differ. The relative free energies of the states in the nonuniform, asymmetric model are drawn from Srinivas et al.\cite{srinivas2013biophysics}.}
    \label{fig:landscapes}
\end{figure}

It is straightforward to derive the first passage time statistics from a Markov chain like Eq.~\ref{eq:1d}. The simplest model is a uniform random walk where all transition rates are equal ($\left\{f_i,r_i \right\}=k$). Such a model can be represented by a free energy landscape shown in Fig.~\ref{fig:landscapes} with troughs separated by equal height barriers. Based on Eq.~\ref{eq:mfpt2}, the mean first passage time ($\tau$) is given by
\begin{equation}
\tau=\frac{1}{2k}n(n+1).
\label{eq:simpletau}
\end{equation}
Using Eq.~\ref{eq:simpletau}, $n=14$, and the measured mean first passage time of \SI{30}{\ms}, we can estimate the single-step migration time ($k^{-1}$) to be \SI{\sim286}{\us}. This estimate is also consistent with the measured histogram of displacement times. If single-step migration occurs more slowly than the time resolution, the histogram of displacement times must exhibit a strong delay or lag in early times (SFig.~\ref{sfig:binning}). However, our measured histograms at 4.4 ms bin width show little or no lag, which points to a single-step migration time much shorter than 4.4 ms.

However, this estimated time of \SI{\sim 330}{\us} per step is likely to be longer than the true value because displacement events faster than the \SI{\sim4}{\ms} exposure time are not included in our measurement. To extract the single-step migration rates in a more accurate, unbiased way despite this missing fraction of events, we fit the analytical solution Eq.~\ref{eq:dwelltime} to all eight histograms with four shared parameters representing rates for A,G,C, and T. This global fitting procedure looks for the best set of rates that describe all eight histograms in the least-squares sense, excluding the missing first bin. It also implies a nonuniform symmetric random walk (Fig.~\ref{fig:landscapes}) where the single-step migration rate depends only on the identity of the base to be displaced. Therefore, each step has the same forward and reverse rates ($f_{i-1}=r_{i}$). The extracted step times for A,C, and T base are \SI{\sim 33}{\us}, \SI{\sim 200}{\us}, and \SI{\sim 250}{\us}, respectively. The step time for the G base did not converge, but the goodness of fit increased with faster values. Thus, we estimate the step time for G to be \SI{\le 33}{\us}. As predicted, these times obtained by fitting histograms in their entirety are all faster than \SI{\sim 330}{\us} obtained from the mean values that omit fast events. 

Our estimated single-step migration rates (\SIrange{0.003}{0.03}{\us^{-1}}) appear to be much slower than the rate of base pair fraying or base flipping ($\SI{\geq1}{\us^{-1}}$)\cite{andreatta2006ultrafast,banavali2013partial,zgarbova2014base,lindahl2017sequence}. Similarly, a previous study by Srivinas et al.\cite{srinivas2013biophysics} also inferred the single-step migration rate to be much slower than the fraying rate. This discrepancy suggests that a single base pair opening event does not always lead to single-step branch migration. As shown in Fig.~\ref{fig:singlestep}, a base pair between the substrate and the incumbent can transiently open and close with rate constants of $k_\mathrm{open}$ and $k_\mathrm{close}$, respectively. While the substrate base is transiently unbound ($\mathrm{S\cdot C^*}$), the invading base can base pair with the substrate base and replace the incumbent base at a rate of $k_\mathrm{replace}$. We can safely assume that $k_\mathrm{close}$ is much faster than $k_\mathrm{open}$ based on the known base pair stability\cite{frank2014fluctuations}. Coarse-grained molecular dynamics simulations\cite{srinivas2013biophysics} show that the branch migration intermediate frequently adopts a coaxially unstacked state where a transiently open incumbent base would be closer to the substrate base than the invading base ($\mathrm{S\cdot C^*}$, Fig.~\ref{fig:singlestep}). Therefore, we reason that $k_\mathrm{close}$ is also much faster than $k_\mathrm{replace}$.  Given $k_\mathrm{close}\gg k_\mathrm{open}, k_\mathrm{replace}$, $\mathrm{S\cdot V}$ will appear at the rate of
\begin{equation}
    k\approx\frac{1}{[\mathrm{S\cdot V}]}\frac{d[\mathrm{S\cdot V}]}{dt}\approx\frac{k_\mathrm{open}}{k_\mathrm{close}} \cdot k_\mathrm{replace}\ll k_\mathrm{open}
    \label{eq:salt}
\end{equation}
Hence, the single-step migration rate $k$ is expected to be much slower than the single base-pair opening rate.

We stress that a symmetric random walk is an oversimplification of strand displacement. As shown in SFig.~\ref{sfig:analyticexperiment}, the symmetric random walk model significantly underrepresents the range of observed displacement times: the fastest observed histogram and the slowest observed histogram are outside the range represented by the fitted curves. Therefore, the observed sequence dependence calls for a more complicated model. We list below several microscopic mechanisms which indicate strand displacement is more properly described as an asymmetric random walk ($f_{i-1}\ne r_i$).

\begin{figure}[th!]
\centering
    \includegraphics[width=12.5cm]{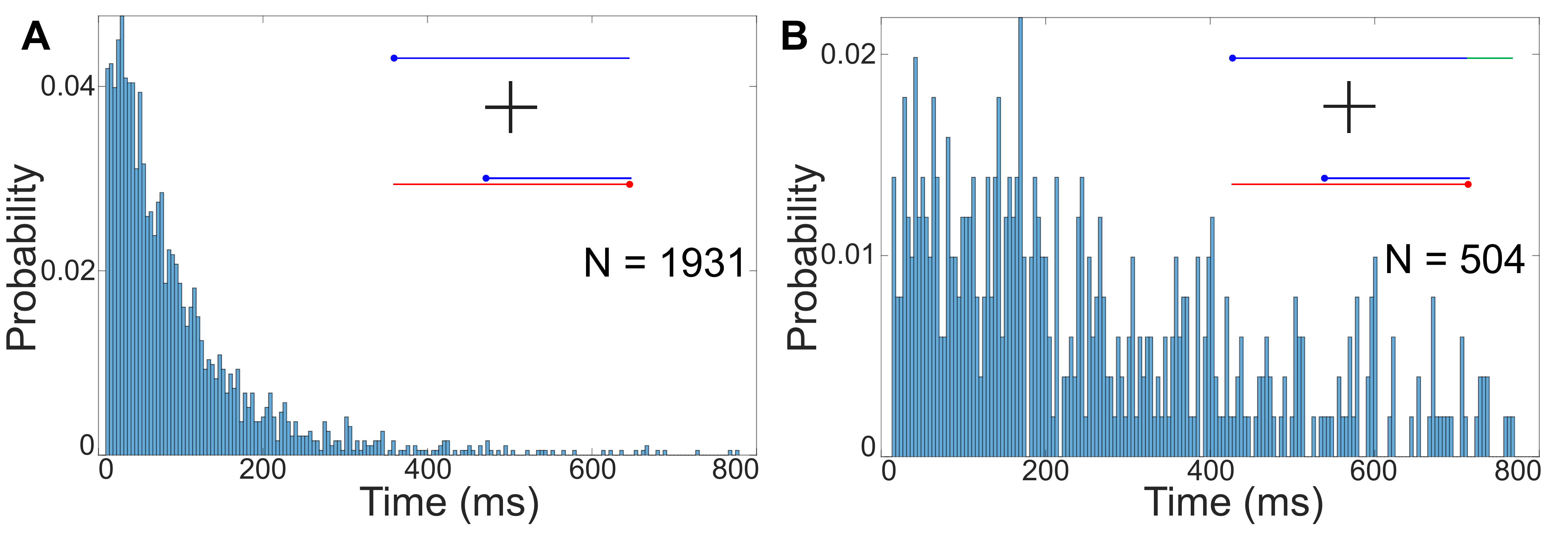}
    \caption{An overhang on the invader strand increases displacement times. (\textbf{A}) A top-invasion experiment with incumbent sequence $5'$-GTCGGAATTTTAAT-$3'$. The distribution of displacement times forms an exponential decay with little lag. After invasion has completed, every base on the invader strand is bound to the substrate. (\textbf{B}) An experiment conducted with a modified invader possessing a 5T overhang on the toehold-distal end (green). These additional bases affect the kinetics of displacement but should not bind to the substrate. The invader overhang increases overall displacement times and introduces a clear lag to the distribution.}
    \label{fig:overhang}
\end{figure}

First, displacement of the first base pair is energetically less favorable than the rest because it creates steric exclusion between dangling bases\cite{srinivas2013biophysics,vsulc2015modelling}. Srinivas et al.\cite{srinivas2013biophysics} measured the thermodynamic penalty for the steric exclusion to be \SI{2.0}{kcal/mol} at \SI{25}{\celsius}, which corresponds to $\sim30$-fold slower $f_0$ than all other rates ($k$). According to Eq.~\ref{eq:mfpt2}, a bias in the first step ($\alpha_1=f_0/r_1$) alters the mean first passage time to
\begin{equation}
    \tau=\frac{1}{2k}n\left(n+\frac{2}{\alpha_1}-1\right).
\end{equation}
With a strong reverse bias ($\alpha_1=1/30$) in the first step, the single-step time ($k^{-1}$) is estimated to be \SI{69}{\us}, faster than our previous estimate of \SI{330}{\us} based on a completely symmetric random walk. \SI{69}{\us}
per step also falls well within the range (\SIrange{53}{103}{\us}) inferred by Srinivas et al.\cite{srinivas2013biophysics}. Second, the stability of a base pair is highly influenced by its nearest neighboring base pair, which would render the base pair opening rate direction-dependent. For example, let us consider an A base in two adjacent branch migration intermediates G$^{\lor}$AC and GA$^{\lor}$C, where $^{\lor}$ refers to the branch point. In G$^{\lor}$AC, A is stacked more closely on C, whereas in GA$^{\lor}$C, A is stacked more closely on G. Therefore, the rate of A flipping out would be different between forward and reverse transitions. Third, the incumbent and the invader base at the branch point carry dangling strands of variable lengths depending on the state. These dangling strands will inevitably affect the diffusion rates of the bases at the branch point. To demonstrate this idea, we performed the fission assay with an invader extended by 5 nucleotides at the $3^\prime$-end. As shown in Fig.~\ref{fig:overhang}, the displacement kinetics become significantly slower even with the same displacement domain. This result is consistent with the idea that a base with a longer dangling strand invades more slowly. As strand displacement progresses, the dangling part of the invader becomes shorter, and the dangling part of the incumbent becomes longer. Hence, the forward rate should become faster ($f_{i-1}<f_{i}$), and the reverse rate slower ($r_{i-1}>r_{i}$). These dangling-strand dependent rates produce asymmetric barriers in the free energy landscape, causing the basins to follow a concave curve (Fig.~\ref{fig:landscapes}, SFig.~\ref{sfig:energylandscape}). Previous oxDNA simulations also predicted a concave free energy landscape\cite{srinivas2013biophysics,vsulc2015modelling}. In the asymmetric random walk model, single-step rates are not only base-dependent but also position-dependent. Determining these rates would require measurements at a much larger scale, which is beyond the scope of this study.         

We assumed that the number of steps is equal to the number of base pairs for modeling purposes. This could raise concern that this number may not accurately reflect the number of actual branch migration steps taken because the incumbent can spontaneously dissociate near the end of migration. In our previous work \cite{broadwater2016effect}, we estimated spontaneous dissociation of a 2-bp incumbent to be \SI{\sim10}{\us} which would be comparable to the branch migration step rate we measured in this study. This means that the last few steps can occur either via branch migration or spontaneous dissociation, and the rate would be dominated by the faster of the two. Regardless, our proposed asymmetric branch migration model, where forward rates become faster, and reverse rates become slower, would effectively account for such an effect.

We made an interesting observation that RNA invasion and DNA invasion occur at very different rates even with the same invader sequence (except for T to U substitution). For the one sequence we tested, RNA invaded faster than DNA (Fig.~\ref{fig:RNA}) from one side but more slowly from the other side. Several factors may contribute to this finding. Structurally, an RNA-DNA hybrid duplex adopts an A-form helix\cite{gyi1998solution,conn1999crystal,de2019dna}, while a DNA-DNA duplex adopts a B-form helix. The thermodynamic stability difference between RNA-DNA and DNA-DNA duplexes depends on the sequence\cite{sugimoto1995thermodynamic}, with purine (AG)-rich substrate favoring RNA-DNA hyrbrid duplexes\cite{huppert2008thermodynamic}. Directional differences in stacking between DNA and RNA are known to persist even in the single stranded form\cite{isaksson2004single} and could contribute to this inversion of side dependence. In a similar vein, a recent study shows that coaxial stacking between an RNA-DNA hybrid duplex and a DNA-DNA homoduplex is stronger when the interhelical junction contains a $5'$ RNA end than when it contains a $3'$ RNA end \cite{cofsky2020crispr}. This effect may partially contribute to the faster top invasion by RNA shown in Fig 5. However, another RNA sequence we tested exhibited faster bottom invasion than top invasion, suggesting that the base sequence is a stronger determinant of the displacement rate than the invasion polarity. RNA invasion of a DNA duplex in particular is a fundamental feature of the CRISPR-Cas system\cite{mulepati2014crystal, singh2016real,hong2019emergent}. R-loop formation appears to be the rate limiting step for DNA cleavage\cite{jeon2018direct,gong2018dna} and is highly sequence dependent\cite{szczelkun2014direct,jeon2018direct,zeng2018initiation}, but proceeds much more slowly ($\sim 1$ second) than the spontaneous displacement rate we measured in this study. It will be thus interesting to investigate whether the sequence dependence is preserved between spontaneous and enzyme-mediated displacement reactions in the future. 

The lack of salt dependence of the measured displacement kinetics was at first surprising to us because salt has a substantial effect on base pairing thermodynamics\cite{santalucia2004thermodynamics}. Experimental measurements of salt-dependent opening and closing rates of a single base pair are scarce, but we can still infer their salt dependence from molecular dynamics study\cite{wang2020salt} and hybridization and dissociation measurements of short oligos\cite{cisse2012rule,dupuis2013single}. These studies show that monovalent cations stabilize base pairing mainly by increasing the rate of base pair closing instead of decreasing the rate of base pair opening. Despite the strong salt dependence of base pair closing ($k_\mathrm{close}$ and $k_\mathrm{replace}$), our proposed three-state model for branch migration (Fig.~\ref{fig:singlestep} and Eq.~\ref{eq:salt}) predicts that salt dependences of $k_\mathrm{close}$ and $k_\mathrm{replace}$ will cancel each other out and render step migration rates, $f$ and $r$, largely salt-independent. 

Even in the case where $f$'s and $r$'s all carry a weak salt-dependence through $k_\mathrm{open}$, we can show that the overall salt dependence of the mean displacement time remains weak. Based on an experimental study\cite{dupuis2013single}, we assume a simple power law dependence of $k_\mathrm{open}$ on $[\mathrm{Na^+}]$  ($k_\mathrm{open}\sim [\mathrm{Na^+}]^\alpha$) so that all $f$'s and $r$'s change by the same factor $c$ upon changing $[\mathrm{Na^+}]$. According to Eq.~\ref{eq:mfpt}, the mean first passage time is equal to the ratio of two matrix determinants
\begin{equation}
    \tau=\frac{\operatorname{det}(\mathbf A')}{\operatorname{det}(\mathbf A)}.
\end{equation}
Since $\operatorname{det}(\mathbf A')\sim c^{n-1}$, and $\operatorname{det}(\mathbf A)\sim c^n$, $\tau\sim c^{-1}$. Hence, the overall displacement of the $n$ base pair domain follows the weak salt dependence of $k_\mathrm{open}$. Either way, we are able to rationalize the weak salt dependence of the mean displacement time (Fig.~\ref{fig:salt}). 

We hope that our results will be beneficial to the field of DNA nanotechnology. Our work has provided sequence-specific branch migration step times that could be used to rationally design sequences with desired kinetics. For example, our results could aid in the design of complex interaction networks between competing reactions with specifically tuned kinetics. Further, our fission assay opens the door to understanding branch migration kinetics in more reaction conditions than we studied here (e.g. buffers, pH, and temperature).

In this study, we assumed that strand displacement proceeds through one-dimensional branch migration, but it is possible that other mechanisms are at play. The invader might invade through the end distal to the toehold when terminal base pairs fray or through internal base pairs that spontaneously open up. Although internal invasion is highly unlikely for the short displacement domain we used here, it would be more probable for longer displacement domains. We also cannot rule out direct swapping between segments of invader and incumbent\cite{paramanathan2014general}, invasion through triplex formation\cite{lee2012kinetics,chen2017parallel}, or concurrent dissociation of a weakly bound incumbent strand\cite{machinek2014programmable,broadwater2016effect}. All these processes can occur in parallel, which makes it difficult to predict the strand displacement rate for any given sequence. In this regard, a future study on a much larger set of displacement domain sequences would help us to attain more accurate phenomenological models for explaining the sequence dependence of strand displacement kinetics.          

\section{Conclusion}
We developed a novel smFRET assay that we call fission in order to study the timing of the unimolecular reaction that occurs during toehold-mediated strand displacement. Our fission assay separates the timescales between the slower toehold formation step and the faster displacement step and enabled us to tally displacement first passage times distributions for 11 separate invasion schemes. We found non-trivial sequence dependence in the distributions, while the mean first passage times varied by an order of magnitude. Further, we highlighted significant differences between the “side” of invasion which suggest the kinetics are not completely determined by base-pair sequence alone. Curiously, we showed that DNA and RNA invaders can behave drastically differently despite having identical sequences (apart from a T$\rightarrow$U substitution). Finally, we demonstrated that displacement times were relatively unchanged over a wide range of salt concentrations. Motivated by these results, we developed a one-dimensional random walk model to estimate single-base displacement times. This model is widely relevant to multistep processes, and we anticipate our analysis to be highly important to an array of biological reactions.

\section{Author Contributions}
DWB and HDK designed research; DWB and AWC performed experiments; DWB, AWC, and HDK contributed analytical tools; DWB and AWC analyzed data; and DWB, AWC, and HDK wrote the manuscript.

\section{Acknowledgements}
The authors thank the current and past members of the Kim laboratory for critical discussions during the research project and helpful comments on the manuscript. This work was supported by National Institutes of Health (R01GM112882) and National Science Foundation (1517507).

$$$$

\bibliography{bibliography}

\clearpage
\newcommand{\beginsupplement}{%
        \setcounter{table}{0}
        \renewcommand{\thetable}{S\arabic{table}}%
        \setcounter{figure}{0}
        \renewcommand{\thefigure}{S\arabic{figure}}%
        \setcounter{equation}{0}
        \renewcommand{\theequation}{S\arabic{equation}}%
        \renewcommand{\figurename}{Supplementary Figure}
        \renewcommand{\tablename}{Supplementary Table}
        \setcounter{page}{1}
}

\beginsupplement
\onecolumngrid
\section*{Supporting Material}

\subsection*{Oligonucleotide sequences used in this study}
\vspace*{-1cm}
\begin{table}[H]
    \centering
    \caption{Invader sequences. The first 8 sequences correspond to the sequences in Figure \ref{fig:meantimes} with top invaders preceding bottom invaders. The RNA sequences are identical to the first two sequences of this table but with a T$\rightarrow$U substitution and correspond to partial duplexes formed from the first two rows of Tables \ref{table:incumbent} and \ref{table:substrate}. The final invader is identical to the first but with a 5T extention on the toehold-distal side, as seen in Fig. \ref{fig:overhang}.}
    \begin{tabular}{ l }
        \specialrule{.1em}{.05em}{.05em} 
        5$'$-/BioTEG/ACCTGGTGTTGTCGGAATTTTAAT-3$'$ \\
        5$'$-ATTAAAATTCCGACAACACCAGGT/BioTEG/-3$'$ \\
        5$'$-/BioTEG/GGTTATTGGGTTGGGTGTGTGAAA-3$'$ \\
        5$'$-TTTCACACACCCAACCCAATAACC/BioTEG/-3$'$ \\
        5$'$-/BioTEG/CAATCAAATAAAACTCTCTCTAAA-3$'$ \\
        5$'$-TTTAGAGAGAGTTTTATTTGATTG/BioTEG/-3$'$ \\
        5$'$-TTTCCTCCTAAAAGACCACCACCT/BioTEG/-3$'$ \\
        5$'$-CACACGTACGAACAAACACCAGGT/BioTEG/-3$'$ \\
        5$'$-/BioTEG/ACCUGGUGUUGUCGGAAUUUUAAU-3$'$ \\
        5$'$-AUUAAAAUUCCGACAACACCAGGU/BioTEG/-3$'$ \\
        5$'$-/BioTEG/ACCTGGTGTTGTCGGAATTTTAATTTTTT-3$'$ \\
        \specialrule{.1em}{.05em}{.05em} 
    \end{tabular}
    \label{table:invader}
\end{table}

\begin{table}[H]
    \centering
    \caption{Incumbent sequences. The order corresponds to Figure \ref{fig:meantimes} with top invaders preceding bottom invaders.}
    \begin{tabular}{ l }
        \specialrule{.1em}{.05em}{.05em} 
        5$'$-GTCGGAATTTTAAT/Cy5/-3$'$ \\
        5$'$-/Cy5/CACACGTACGAACA-3$'$ \\
        5$'$-TTGGGTGTGTGAAA/Cy5/-3$'$ \\
        5$'$-/Cy5/TTTCACACACCCAA-3$'$ \\
        5$'$-AAACTCTCTCTAAA/Cy5/-3$'$ \\
        5$'$-CTTTTAGGAGGAAA/Cy5/-3$'$ \\
        5$'$-/Cy5/GACGACTATAGTTC-3$'$ \\
        5$'$-TGTTCGTACGTGTG/Cy5/-3$'$ \\
        \specialrule{.1em}{.05em}{.05em} 
    \end{tabular}
    \label{table:incumbent}
\end{table}

\begin{table}[H]
    \centering
    \caption{Substrate sequences. The order corresponds to Figure \ref{fig:meantimes} with top invaders preceding bottom invaders.}
    \begin{tabular}{ l } 
    	\specialrule{.1em}{.05em}{.05em} 
        5$'$-CA/Cy3/ATTAAAATTCCGACAACACCAGGT-3$'$ \\
        5$'$-ACCTGGTGTTGTCGGAATTTTAAT/Cy3/AC-3$'$ \\
		5$'$-CA/Cy3/TTTCACACACCCAACCCAATAACC-3$'$ \\
		5$'$-GGTTATTGGGTTGGGTGTGTGAAA/Cy3/AC-3$'$ \\
        5$'$-CA/Cy3/TTTAGAGAGAGTTTTATTTGATTG-3$'$ \\
        5$'$-CAATCAAATAAAACTCTCTCTAAA/Cy3/AC-3$'$ \\
        5$'$-AGGTGGTGGTCTTTTAGGAGGAAA/Cy3/AC-3$'$ \\
		5$'$-ACCTGGTGTTTGTTCGTACGTGTG/Cy3/AC-3$'$ \\
        \specialrule{.1em}{.05em}{.05em} 
    \end{tabular}
    \label{table:substrate}
\end{table}

\begin{figure}[!th]
\begin{minipage}[c][\textheight]{\textwidth}
\includegraphics[width=10cm]{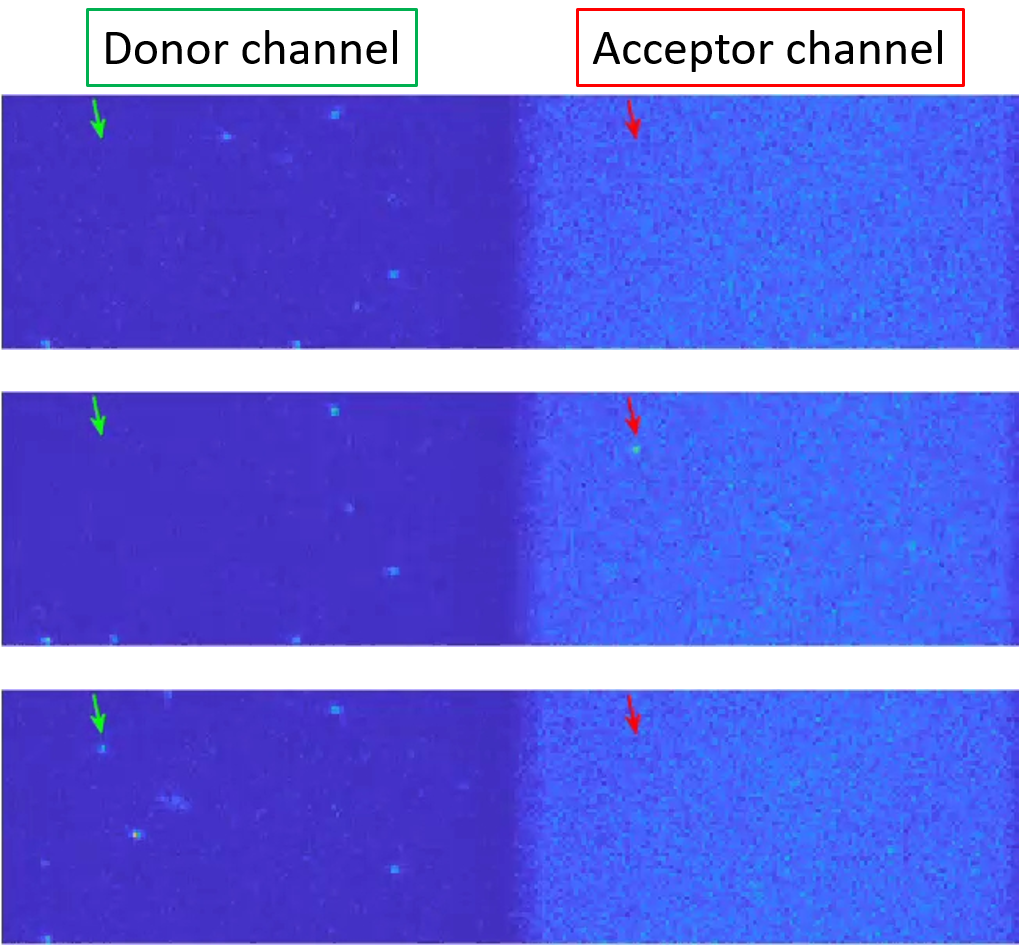}
\caption{Unprocessed single frame data acquired by the EMCCD with an exposure time of 3.96 ms. The green and red arrows track a single displacement event in the donor and acceptor channels, respectively. The top image shows no fluorescence signal. The middle image shows a fluorescent spot in the acceptor channel due to binding of the partial duplex to the invader. The bottom image shows a fluorescent spot in the donor channel due to the dissociation of the incumbent strand.}
\label{sfig:rawdata}
\end{minipage}
\end{figure}

\begin{figure}[!th]
\begin{minipage}[c][\textheight]{\textwidth}
\includegraphics[width=17cm]{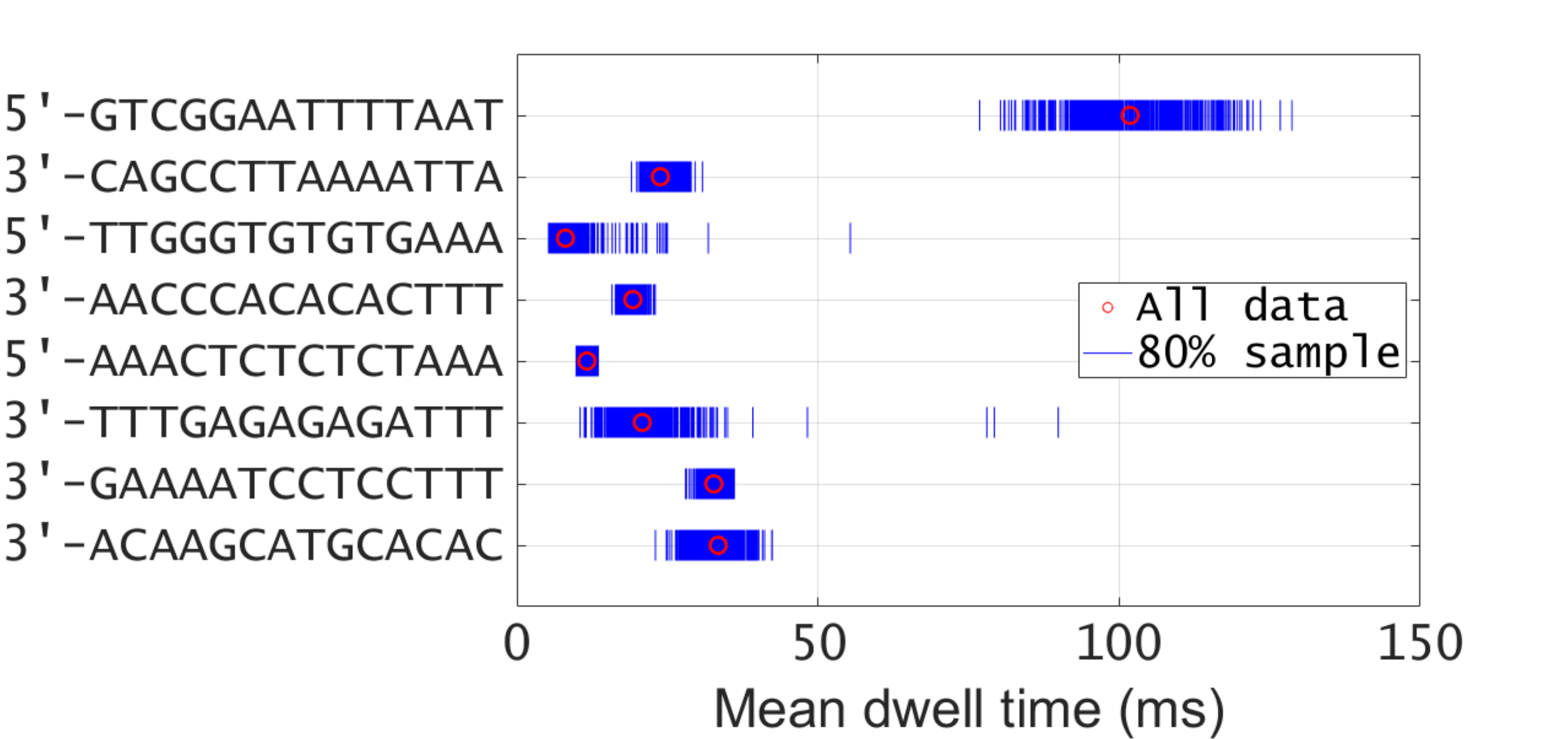}
\caption{A representation of the variance of mean displacement times. The mean displacement time is shown in red for one series of experiments using each of the eight sequences employed in this study. Each blue line represents the mean of a random sample of the data points within the experiment. Each sample is 80\% the size of the total data set and allows for replacement between each selected point; 500 such samples have been taken for each sequence. Before each mean is calculated, the longest 5\% of dwell times are discarded. This result demonstrates that the separation in mean dwell times between different sequences is not merely the result of insufficiently robust statistics.}
\label{sfig:mean sampling}
\end{minipage}
\end{figure}

\begin{figure}[!th]
\begin{minipage}[c][\textheight]{\textwidth}
\includegraphics[width=17cm]{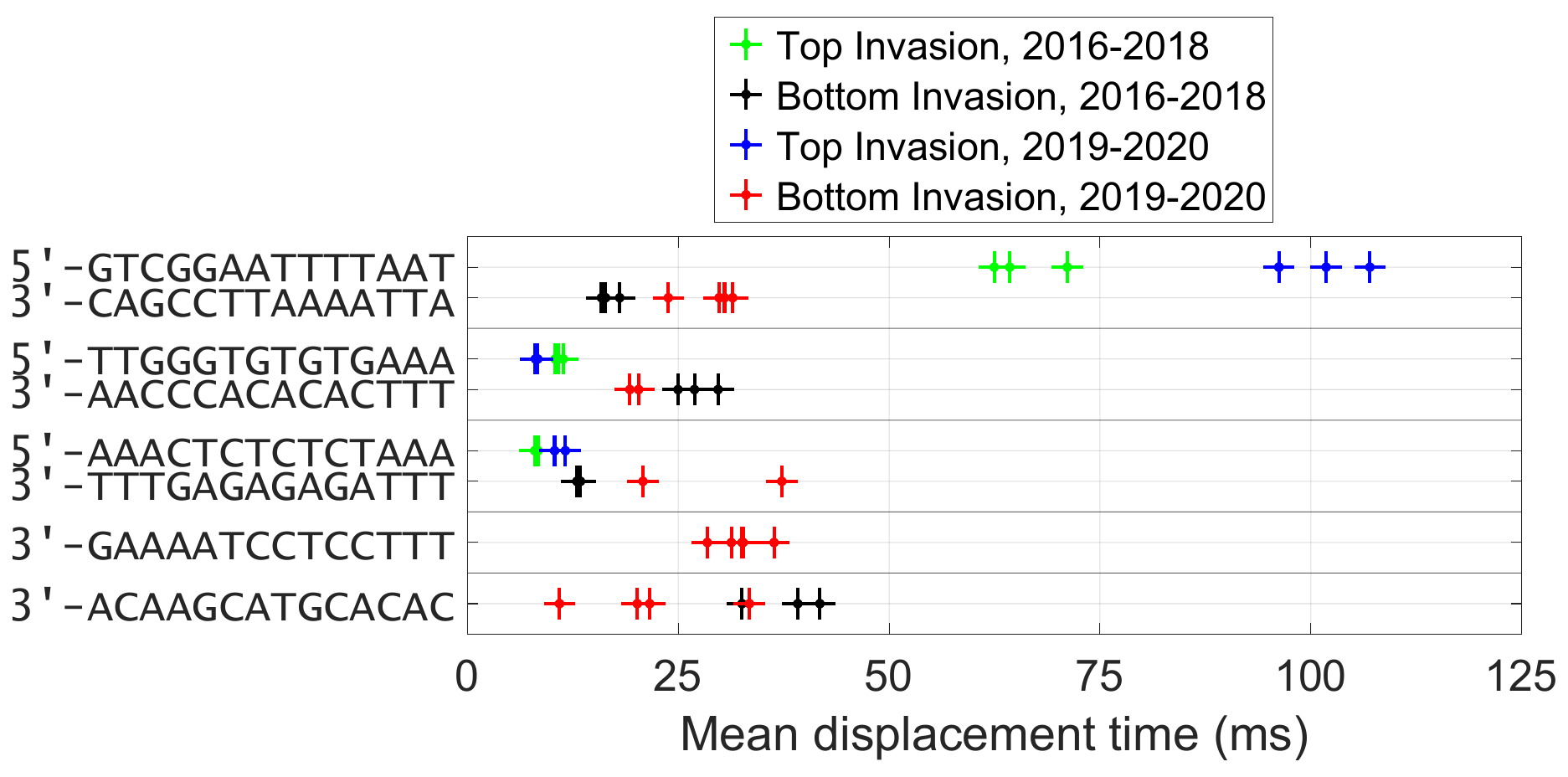}
\caption{A comparison between mean dwell times calculated over the course of four years. Two different graduate students worked on this project. Student 1 collected data between 2016 and 2018, while Student 2 collected data from 2019 to 2020. The sequences of the corresponding incumbent strands are labelled on the left. }
\label{sfig:means old new}
\end{minipage}
\end{figure}

\begin{figure}[!th]
\begin{minipage}[c][\textheight]{\textwidth}
\includegraphics[width=17cm]{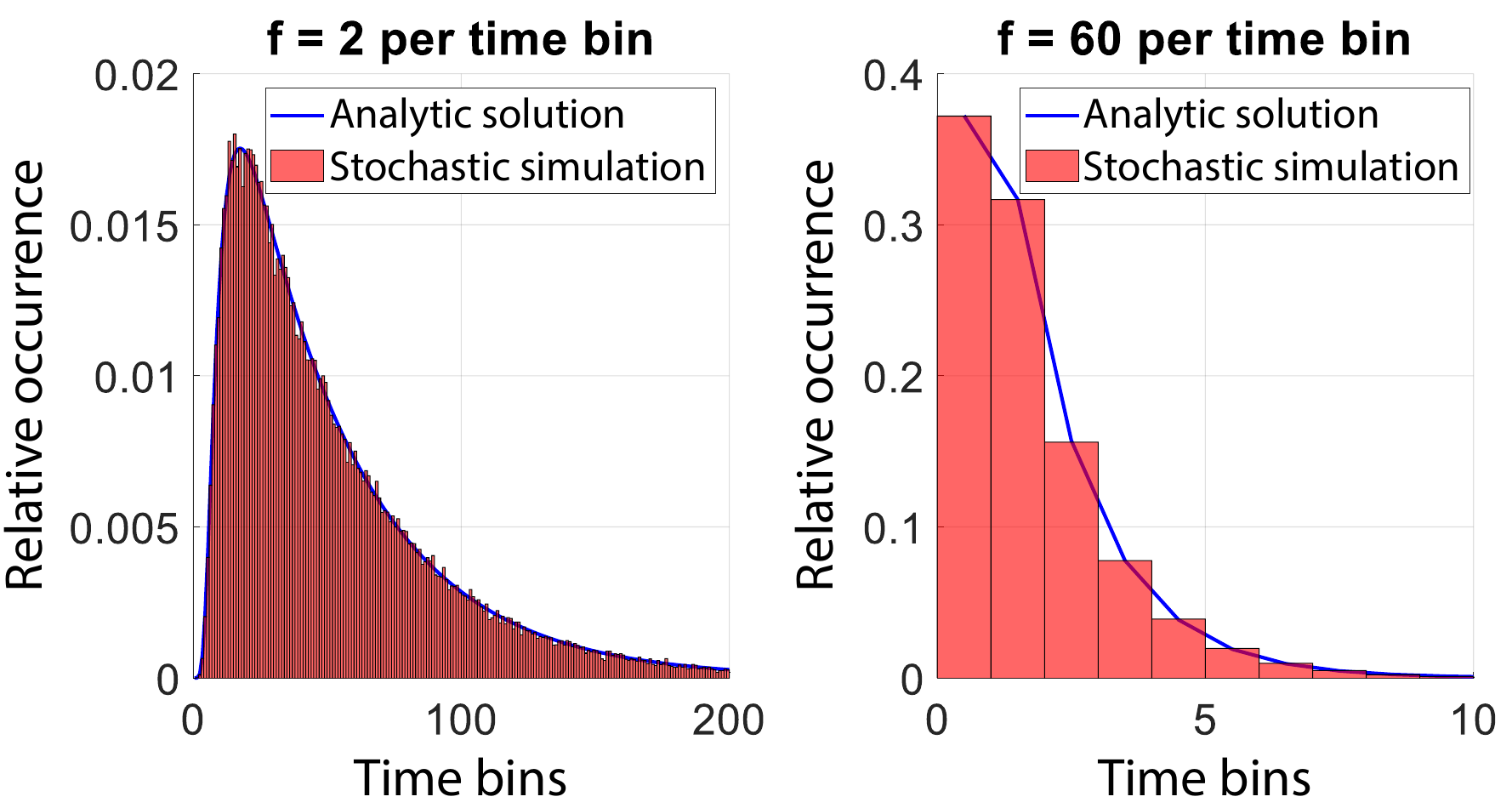}
\caption{Dwell-time histogram vs. bin width. Both plots show the normalized histograms of first passage times of a homogeneous random walker on a lattice with 14 sites. The red histograms are produced from a stochastic simulation (n = 10000), while the blue lines represent the analytic solution given by Eq. \ref{eq:dwelltime}. A key feature of multi-step processes is a noticeable lag in the first passage time distribution (left). The reason for this is that some time must pass for a random walker to reach the end if there are multiple steps. However, if the binning time is large compared to the single step rate, then the lag can disappear (right). Further, for large binning times, the mean first passage time (measured in bins) must equal about half the number of lattice sites which implies the distribution will have a very short tail.}
\label{sfig:binning}
\end{minipage}
\end{figure}

\begin{figure}[!th]
\begin{minipage}[c][\textheight]{\textwidth}
\includegraphics[width=17cm]{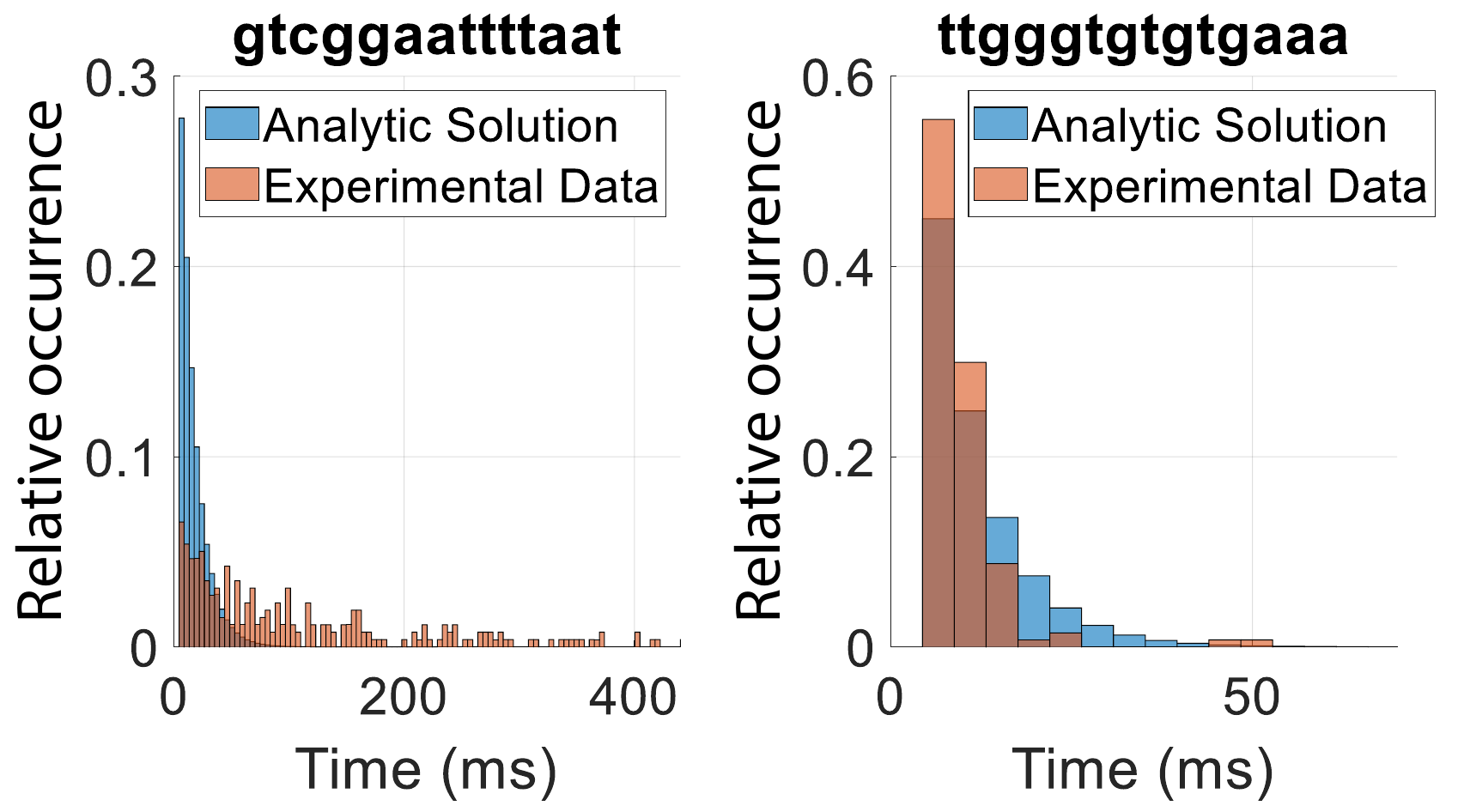}
\caption{Analytic solutions compared to experimental results. Both histograms show comparisons between experimentally obtained distributions of first passage times (orange histograms) and the solution yielded by Eq. \ref{eq:dwelltime} using the best-fit A, C, and T rates (blue histograms). The slowest-displacing sequence (left) and the fastest (right) are shown here. Both plots are labelled with the sequence of the invasion domain in the $5'$ to $3'$ direction. }
\label{sfig:analyticexperiment}
\end{minipage}
\end{figure}

\newpage
\subsection*{Asymmetric random walk model}
\begin{figure}[!th]
\begin{minipage}[c][\textheight]{\textwidth}
\includegraphics[width=17cm]{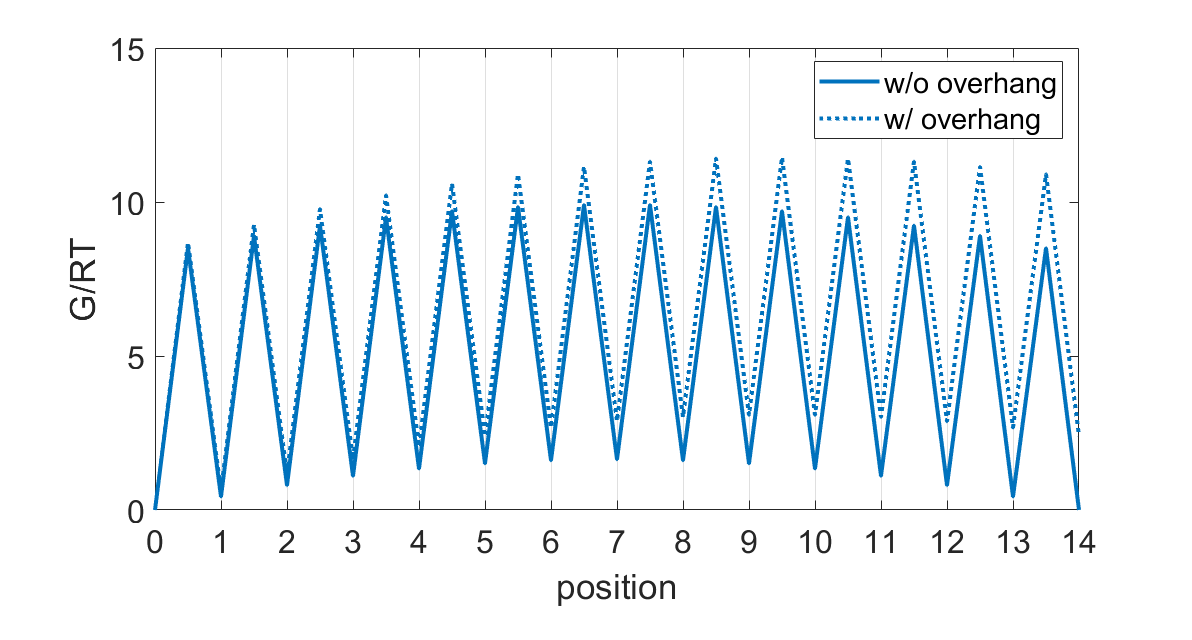}
\caption{Free energy landscape for asymmetric random walk. As strand displacement progresses, the length of the invader becomes shorter while the length of the incumbent becomes longer. Therefore, we expect the forward rate to become faster, and the reverse rate to become slower with the position along the free energy landscape. A free energy landscape that is consistent with this kinetic model is represented by the solid line. Extending the invader by an overhang of five bases causes the landscape to tilt, as shown by the dotted line.}
\label{sfig:energylandscape}
\end{minipage}
\end{figure}

We attempt to expand the random walk model to explain the 3.3-fold slower mean strand displacement time with the longer invader. This result suggests that the forward transition barrier would decrease with position as the overhang length of the invader becomes shorter. Likewise, the reverse transition barrier should increase with position. Although the exact position dependence of the barrier height is not known, we assume a simple analytical form (Hill equation or exponential function) for forward and reverse barrier heights (SEqs.~\ref{seq:hillf} and ~\ref{seq:hillr}).
\begin{equation}
\Delta G^*_f(i) = \frac{8.5 RT}{1+i/K}
\label{seq:hillf}
\end{equation}
\begin{equation}
\Delta G^*_r(i) = \frac{8.5 RT}{1+(14-i)/K}
\label{seq:hillr}
\end{equation}
Here, $i$ represents the position along the energy landscape ($i\in\{0,1,...,14\}$), the numerator comes from a previous estimate \cite{srinivas2013biophysics}, and the factor $K$ in the denominator determines the concavity of the landscape. For example, by substituting $K \approx 240$, we obtain a concave energy landscape as shown in SFig.~\ref{sfig:energylandscape}, whose midpoint (position 7) of the landscape is about \SI{1}{kcal/mol} higher than the end states (position 0 or 14). This concavity is similar to that seen in the landscape constructed from simulations \cite{srinivas2013biophysics}. If the invader is lengthened by 5 extra bases, all forward barriers become higher:
\begin{equation}
\Delta G^*_f(i) = \frac{8.5 RT}{1+(i-5)/K}
\label{seq:hill-long}
\end{equation}
whereas the reverse barriers are unchanged (SEq.~\ref{seq:hill-long}). Hence, the energy landscape becomes tilted (dotted line, SFig.~\ref{sfig:energylandscape}), and the corresponding mean first passage time is predicted to be slower by 3.3-fold based on Eq.~\ref{eq:mfpt} of the main text. This crude exercise shows that our observed overhang effect is semi-quantitatively consistent with a simple asymmetric random walk model.

\end{document}